\documentclass[runningheads]{llncs}
\usepackage[T1]{fontenc}
\usepackage{amsmath}
\usepackage{amssymb}
\usepackage{booktabs}
\usepackage{multirow}
\usepackage{multicol}
\usepackage{longtable}
\usepackage{graphicx,verbatim}
\usepackage{marvosym}
\newcommand{\bsigma}{\boldsymbol{\Sigma}}
\newcommand{\bmu}{\boldsymbol{\mu}}
\newcommand{\bx}{\boldsymbol{x}}
\newcommand{\br}{\boldsymbol{R}}
\newcommand{\bs}{\boldsymbol{S}}
\newcommand{\bu}{\boldsymbol{u}}
\usepackage{hyperref}
\usepackage{color}

\begin{document}
\title{DIGS: Dynamic CBCT Reconstruction using Deformation-Informed 4D Gaussian Splatting and a Low-Rank Free-Form Deformation Model}
\titlerunning{DIGS for dynamic CBCT}

\author{Yuliang Huang\inst{1}\textsuperscript{(\Letter)} \and
Imraj Singh\inst{2} \and
Thomas Joyce\inst{3} \and
Kris Thielemans\inst{1,4} \and
Jamie R. McClelland\inst{1}}

\authorrunning{Y. Huang et al.}
%
\institute{
UCL Hawkes Institute, Department of Medical Physics and Biomedical Engineering, University College London, United Kingdom \email{yuliang.huang.21@ucl.ac.uk} \and
Department of Computer Science, University College London, United Kingdom \and
Elekta, AB, Stockholm, Sweden \and
Institute of Nuclear Medicine, University College London, United Kingdom 
}
    
\maketitle              
\begin{abstract}
3D Cone-Beam CT (CBCT) is widely used in radiotherapy but suffers from motion artifacts due to breathing. A common clinical approach mitigates this by sorting projections into respiratory phases and reconstructing images per phase, but this does not account for breathing variability. Dynamic CBCT instead reconstructs images at each projection, capturing continuous motion without phase sorting.
Recent advancements in 4D Gaussian Splatting (4DGS) offer powerful tools for modeling dynamic scenes, yet their application to dynamic CBCT remains underexplored. Existing 4DGS methods, such as HexPlane, use implicit motion representations, which are computationally expensive. While explicit low-rank motion models have been proposed, they lack spatial regularization, leading to inconsistencies in Gaussian motion. To address these limitations, we introduce a free-form deformation (FFD)-based spatial basis function and a deformation-informed framework that enforces consistency by coupling the temporal evolution of Gaussian's mean position, scale, and rotation under a unified deformation field.
We evaluate our approach on six CBCT datasets, demonstrating superior image quality with a 6× speedup over HexPlane. These results highlight the potential of deformation-informed 4DGS for efficient, motion-compensated CBCT reconstruction. The code is available at \url{https://github.com/Yuliang-Huang/DIGS}.


\end{abstract}

\section{Introduction}
Cone-beam CT (CBCT) is widely used in image-guided radiotherapy \cite{de2013image} but is prone to artifacts caused by patients' involuntary motion, such as respiration \cite{sweeney2012accuracy}. A common clinical approach to address this issue is 4DCBCT, which bins projections into different respiration phases and reconstructs a separate image for each phase \cite{sonke2005respiratory}. However, 4DCBCT requires extra acquisition time and radiation dose, and is susceptible to artifacts caused by phase sorting errors \cite{leng2008streaking}. More critically, it assumes that patients' breathing motion is regular, which is often not the case \cite{huang2024resolving}. To account for irregular motion, dynamic CBCT reconstruction aims to estimate both the motion and the image on a per-projection basis \cite{huang2024surrogate,jailin2021projection}, where a fast motion-compensated reconstruction algorithm is needed to satisfy the demands of clinical feasibility.

Traditional reconstruction methods typically rely on a discrete voxel grid representation. In recent years, implicit neural representations (INR) have gained attention for their inherent continuity and resolution independence \cite{zha2022naf}. However, INR suffers from low computation efficiency because each query point needs to be evaluated by a multilayer perceptron (MLP). More recently, 3D Gaussian Splatting (3DGS) has emerged as a promising alternative \cite{kerbl20233d}, representing the image volume as a Gaussian point cloud and enabling fast rendering through efficient splatting techniques \cite{zwicker2002ewa}. While 3DGS has been applied to static CBCT \cite{r2_gaussian}, applying 4D Gaussian Splatting (4DGS) for dynamic CBCT reconstruction remains underexplored, which is the primary objective of this paper. 

To the best of our knowledge, only one concurrent work \cite{fu2025spatiotemporal} applies 4DGS for 4DCBCT reconstruction, which still requires phase sorting. In this work, we introduce several methodological innovations upon existing approaches: 

\textbf{1.} Most current 4DGS methods, including those in non-medical applications, model deformation with implicit representations, such as Neural Field \cite{park2021nerfies} or HexPlane \cite{cao2023hexplane}, which incurs significant computational overhead due to MLP inference. In contrast, we propose a more efficient explicit motion representation based on Free-Form Deformation (FFD).

\begin{figure}[!t]
    \centering
    \includegraphics[width=\textwidth]{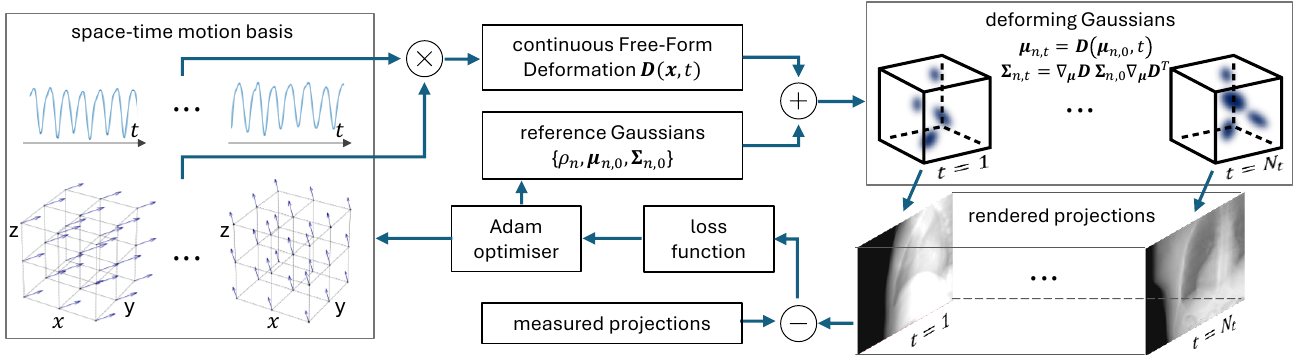}
    \caption{Schematic illustration of the proposed method. A continuous free-form deformation (FFD) is constructed from a set of learnable spatial and temporal basis, which is used to warp the reference Gaussians though a deformation-informed (DI) framework. Projections rendered from the dynamic Gaussians are compared with the measured projections and the optimiser updates the reference Gaussians and the motion basis to minimize their difference. }
    \label{fig1}
\end{figure}

\textbf{2.} Existing 4DGS methods typically fit motion by independently modifying the Gaussian kernel's mean positions, scales and rotations. Although these methods can produce plausible dynamic images, they lack an interpretable spatial correspondence between time frames. This is a critical limitation in applications such as tumor tracking in radiotherapy, where the voxel-wise deformation vector field (DVF) itself is as important as the reconstructed images. To address this, we propose a deformation-informed (DI) framework that consistently updates all Gaussian attributes according to a unified DVF. Figure~\ref{fig1} provides a schematic of the proposed method.

\section{Related Work}
\subsection{Explicit Motion Representation}
Several studies have used explicit representation for motion fields in 4DGS. For example, Gaussian-flow \cite{lin2024gaussian} parameterizes the motion of each Gaussian as polynomial curves and Fourier series. Other studies \cite{kratimenos2024dynmf,cai2024gaussian} factorize the motion field into a linear combination of a few learnable basis. Yang et al. \cite{yang2024deform3dgs} also use a low-rank motion model but constrain the temporal basis to be Gaussian functions. These methods fit the weights of motion basis for each Gaussian independently, without imposing any smoothness constraint on the motion of adjacent Gaussians. In comparison, our proposed method interpolates the basis weights from a free-form deformation field \cite{modat2010fast,rueckert1999nonrigid} that enforces adjacent Gaussians to move consistently with each other.

\subsection{Applying Deformation Fields to Gaussian Kernels}
Most 4DGS methods learn the temporal changes of the Gaussian's mean position, scale and rotation independently. In contrast, inspired by continuum mechanics \cite{bonet1997nonlinear}, physics-based 3DGS simulates dynamic scenes by explicitly applying continuous DVF to Gaussians, by deriving the transformation of scale and rotation from the Jacobian of mean position deformation \cite{xie2024physgaussian,lin2024phys4dgen,feng2024gaussian}. While these previous works focus on applying synthetic deformation to static objects, we are the first to apply a physics-based DI framework for handling motion in image reconstruction problems.

\section{Methodology}
\subsection{Preliminary: 3DGS for CBCT reconstruction}
Given a set of measured projections $\{P_t\}_{t=1,...,N_t}$ for $N_t$ time points, the problem of CBCT reconstruction can be formulated as an inverse problem:
\begin{equation}\label{eq:obj}
    \min_{\phi_{\bx}} \sum_{t=1}^{N_t}L( \mathbb{F}(\phi_{\bx}),P_t)
\end{equation}
where $\mathbb{F}$ is the X-ray transform, $\phi_{\bx}$ is the attenuation field, and $L$ is the dissimilarity between re-projection and measured data.

Defining a set of Gaussian kernels $\{G_n\}_{n=1,...,N_g}$ as 
\begin{equation}
    G_n(\bx|\rho_n,\bmu_n,\bsigma_n)=\rho_n e^{-\frac{1}{2}(\bx-\bmu_n)^T\bsigma_n^{-1}(\bx-\bmu_n)}
\end{equation}
where $\rho_n$ is the density, $\boldsymbol{\mu}_n$ is the mean position, $\bsigma_n=\br_n \bs_n \bs^T_n \br^T_n$ is the covariance matrix that can be decomposed into scale $\bs_n$ and rotation $\br_n$. $\bs_n$ is a diagonal matrix and $\br_n$ is an orthonormal matrix parameterized by quaternion. The attenuation field can be represented as the superposition of these Gaussian kernels:
\begin{equation}
    \phi_{\bx}=\sum_{n=1}^{N_g}G_n(\bx|\rho_n,\bmu_n,\bsigma_n)
\end{equation}

Cone-beam projection of a 3D Gaussian onto a 2D plane can be approximated by a 2D Gaussian:
\begin{equation}
    \tilde{G}_n=\tilde{\rho}_n e^{-\frac{1}{2}(\tilde{\bx}-\tilde{\bmu}_n)^T\tilde{\bsigma}_n^{-1}(\tilde{\bx}-\tilde{\bmu}_n)}
\end{equation}
The tilde over the letters refers to the 2D counterpart of the variable, i.e. $\tilde{\bx}$ and $\tilde{\bmu}_n$ are the projections of $\bx$ and $\bmu_n$. $\tilde{\bsigma}_n$ can be obtained by dropping the third row and column of $\boldsymbol{J}_n\boldsymbol{W} \bsigma_n \boldsymbol{W}^T\boldsymbol{J}_n^T$ where $\boldsymbol{W}$ is the projection-angle dependent viewing transform and $\boldsymbol{J}_n$ is the local Jacobian of the perspective transform. $\tilde{\rho}_n$ is $\rho_n$ scaled by the integration bias factor proposed in \cite{r2_gaussian}. The X-ray transform of $\phi_{\bx}$ can then be approximated by the superposition of the 2D Gaussians, i.e. $\mathbb{F}(\phi_{\bx}) \simeq \sum_{n=1}^{N_g}\tilde{G}_n(\tilde{\bx})$. A gradient-based method is then used to optimize Eqn(\ref{eq:obj}) to fit the attributes of each Gaussian, i.e. $\{\rho_n,\bmu_n,\boldsymbol{R}_n, \boldsymbol{S}_n\}_{n=1,...,N_g}$.

\subsection{Deformation-Informed 4DGS}
4DGS extends 3DGS to dynamic scenes by allowing the attributes of the Gaussians to vary over time. Many 4DGS methods use separate models for estimating the change of each attribute, i.e.
\begin{equation}
    \boldsymbol{v}_{n,t} = \boldsymbol{v}_{n,0} + f_{\boldsymbol{v}_n}(t)
\end{equation}
where $\boldsymbol{v}_{n,t}$ refers to the values of any attribute in $\{\rho_n,\bmu_n,\boldsymbol{R}_n, \boldsymbol{S}_n\}_{n=1,...,N_g}$ at time $t$, and $f_{\boldsymbol{v}_n}(t)$ is an attribute-specific function that gives the change of $\boldsymbol{v}_{n,t}$ against $\boldsymbol{v}_{i,0}$ at time $t$.

From this, the dynamic images can be generated by 
\begin{equation}
    \phi_{\bx}(t)=\sum_{n=1}^{N_g}G_n(\bx|\rho_{n,t},\bmu_{n,t},\bsigma_{n,t}), \ t=1,...,N_t.
\end{equation}
However, if changes of Gaussian parameters are modeled independently, it is impossible to derive a continuous DVF that is consistent with the rendered dynamic images, i.e. one cannot define a transformation that maps a voxel-grid representation of one dynamic image to another. To see this, consider two initially overlapping Gaussian that move apart over time, there is no spatial transformation that can map the overlapping Gaussians to the two separate Gaussians.

Instead of allowing each attribute to evolve independently, we follow a physics-based 3DGS approach \cite{xie2024physgaussian} and warp the Gaussian kernels using a unified DVF. More specifically, the mean positions are shifted directly with the DVF $D$, i.e.
\begin{equation}
    \bmu_{n,t}=D(\bmu_{n,0},t)
\end{equation}
and the covariance matrices are derived by approximating $D$ by a locally linear deformation, i.e.
\begin{equation}
    \bsigma_{n,t} \simeq \boldsymbol{K}\bsigma_{n,0}\boldsymbol{K}^T, \ \boldsymbol{K}=\nabla_{\bmu}D(\bmu_{n,0},t)
\end{equation}
and $\rho_n$ remains constant over time. This ensures that all attributes of the Gaussian kernels are updated consistently by the same DVF. 

\subsection{Explicit Low-Rank B-Spline Motion Model}
The DVF can be decomposed into a linear combination of a few spatial and temporal basis functions, $\bu_r(\bx)$ and $\omega_r(t)$, i.e.
\begin{equation}
    D(\bx,t) = \bx + \sum_{r=1}^{N_r}\omega_r(t)\bu_r(\bx)
\end{equation}
where $N_r$ is the number of basis functions, much smaller than $N_g$ and $N_t$.

Each spatial basis $\bu_{r}(\bx)$ is explicitly parameterized by a set of control points arranged on a uniform $N_x\times N_y\times N_z$ lattice with predefined spacing, and each control point stores a displacement vector $\boldsymbol{d}^{r}_{ijk}$, where $i\in\{1,...,N_x\},j\in\{1,...,N_y\},k\in\{1,...,N_z\}, r\in\{1,...,N_r\}$. 
Let $\bx$ be a point in world space and $\boldsymbol{p}=(p_1,p_2,p_3)$ its corresponding coordinates in the lattice index space. $\bu_r(\bx)$ is interpolated from the $4\times4\times4$ control points surrounding $\bx$:
\begin{equation}
    \bu_r(\bx) = \sum_{l=0}^{3}\sum_{m=0}^{3}\sum_{n=0}^{3}
    B_l(p_1-\lfloor p_1 \rfloor)B_m(p_2-\lfloor p_2 \rfloor)B_n(p_3-\lfloor p_3 \rfloor)\boldsymbol{d}^{r}_{i,j,k}
\end{equation}
where $i=\lfloor p_1 \rfloor-1+l$, $j=\lfloor p_2 \rfloor-1+m$, $k=\lfloor p_3 \rfloor-1+n$,  $\lfloor \cdot \rfloor$ is the floor function and $B_l$ is the $l^{th}$ cubic B-spline basis function \cite{rueckert1999nonrigid}.
The Jacobian is given by
\begin{equation}
    K(\bx,t) = \boldsymbol{I} + \sum_{r=1}^{N_r}\omega_r(t)
\nabla\bu_r(\bx)
\end{equation} 
$\omega_r(t)$ is also parameterized by a 1D B-Spline. A customized CUDA kernel is implemented to efficiently perform differentiable cubic B-spline interpolation.

\section{Experiment Settings}
\subsection{Datasets}
\subsubsection{XCAT Simulation} The 4DXCAT software \cite{segars20104d} was used to generate digital phantom images with dimension of $375\times375\times345$ and resolution of 1 mm as well as simulated motion controlled by breathing traces. The raw DVFs were post-processed by cidX software \cite{eiben2020consistent} to avoid folding and preserve sliding motion.  RTK \cite{rit2014reconstruction} was then used to generate 310 projections in a full circle with dimension of 512$\times$512 and resolution of 0.8 mm. The detector was offset by 116 mm. Distances from source to isocenter and source to detector are 1000 mm and 1536 mm respectively. This is the same as the acquisition geometry of a real CBCT scan. Poisson noise ($\lambda=10^8$) and Gaussian noise ($\sigma=4$) was added to the projections to simulate quantum and electronic noise. Two sets of breathing traces were measured from cine MR images of real patients and used as input to XCAT to create two datasets, denoted as case 1 and 2 respectively.
\subsubsection{Simulation from 4DCT patient images} Simulated data were generated using clinical 4DCT images from four patients in the 4DLung dataset \cite{balik2013evaluation}. All phases were registered to the end-exhalation phase using NiftyReg \cite{modat2010fast}, with default parameters except for the use of sliding registration \cite{eiben2018statistical}. Principal Component Analysis (PCA) was then applied to the stack of deformation fields, preserving the mean deformation and the two dominant principal components. To simulate 20 breathing cycles, the fitted principal component weights were replicated while introducing a slow drift. Additionally, the motion magnitude of each breathing cycle was scaled by a random factor between 0.8 and 1.2 to simulate variability in breathing motion. 310 time points were then uniformly sampled from the 20 breathing cycles to generate CBCT projections, following the same procedure as for XCAT data. The four simulations are denoted as case 3-6 respectively. 

\subsection{Implementation Details}
The Gaussian point cloud is initialized by sampling 80K points from conjugate gradient least square reconstruction, in line with \cite{fu2025spatiotemporal} for fair comparison. The B-Spline control points have a spacing of every 4 time points and 8 voxels. $N_r$ is set to 2. L2 loss in projection space is used for training.
The reference Gaussians and the space-time B-Spline grids are jointly updated with Adam optimizer for 50K iterations. Initial learning rates for the space and time B-Spline grids are set to $10^{-4}$ and $10^{-2}$ respectively, and decay linearly to $10^{-5}$ and $10^{-3}$ by the final iteration. The gradient threshold for density control is $5\times10^{-8}$. All the other hyper-parameters are the same as in \cite{r2_gaussian}.

\subsection{Evaluation} 
Our proposed method is compared with two other methods. SuPReMo \cite{huang2024surrogate} fits a linear model between the DVF and low-dimensional surrogate signals extracted from projections, and uses motion compensated FDK \cite{rit2009fly} as the reconstruction algorithm. HexPlane-based 4DGS method \cite{fu2025spatiotemporal} encodes spatial-temporal features with six 2D grids which are then fed into multi-head MLP to obtain the time-varying attributes per Gaussian kernel. The original method has only been applied on phase-sorted data, where projections of the same phase are treated as the same time-point. To compare with the proposed method, each projection is treated as an individual time-point.

We also perform two ablation studies. To investigate the impact of ensuring that all Gaussian attributes change consistently with the underlying FFD, Ablation Study 1 allows each Gaussian attribute to vary independently over time. Building on this, Ablation Study 2 further removes the FFD representation entirely; that is, the spatial basis is no longer interpolated from the control point grid but instead stored as a separate attribute for each Gaussian. 

All the experiments are run on a single RTX3090 GPU.  The results are evaluated by the average Root-Mean-Squared-Error (RMSE) and Peak-Signal-Noise-Ratio (PSNR) between the estimated dynamic images at each time-point and the ground-truth. The ground truth is created by warping reference images with the simulated DVFs. All evaluations are within the reconstruction FOV.

\section{Results}

\begin{table}[!b]
\centering
\caption{ Performance of proposed method on simulation data in comparison with SuPReMo~\cite{huang2024surrogate} and HexPlane~\cite{fu2025spatiotemporal} as well as ablation studies  excluding deformation-informed (DI) framework and Free-Form Deformation (FFD). Best results are in bold.}
\label{tab1}
\setlength{\tabcolsep}{0.5pt}
\begin{tabular}{@{}cccccccccc@{}} 
\toprule
\multirow{2}{*}{Case} & \multicolumn{3}{c}{PSNR (dB) $\uparrow$} & \multicolumn{3}{c}{RMSE ($10^{-3}$ $\mathrm{mm}^{-1})$ $\downarrow$} & \multicolumn{3}{c}{Time $\downarrow$} \\
\cmidrule(lr){2-4} \cmidrule(lr){5-7} \cmidrule(lr){8-10} 
                        & SuPReMo & HexPlane & Ours & SuPReMo & HexPlane & Ours & SuPReMo & HexPlane & Ours \\ 
\midrule
1 & 18.49 & 21.98 & \textbf{25.46} & 1.63 & 1.91 & \textbf{1.28} & \textbf{47s} & 1h31m & 16m14s  \\
2 & 18.26 & 21.15 & \textbf{21.78} & \textbf{1.88} & 2.12 & 1.97 & \textbf{1m} & 1h27m & 15m48s \\
3 & 21.46 & 26.82 & \textbf{29.48} & 5.08 & 2.78 & \textbf{2.02} & \textbf{41s} & 1h43m & 14m34s \\
4 & 24.79 & 28.48 & \textbf{30.94} & 3.46 & 2.29 & \textbf{1.71} & \textbf{1m2s} & 1h49m & 14m1s \\
5 & 24.66 & 29.35 & \textbf{31.66} & 3.51 & 2.10 & \textbf{1.58} & \textbf{45s} & 1h38m & 16m14s \\
6 & 20.79 & 27.36 & \textbf{29.49} & 5.39 & 2.68 & \textbf{1.99} & \textbf{48s} &  1h50m & 16m26s \\
Mean & 21.41 & 25.86 & \textbf{28.14} & 3.49 & 2.31 & \textbf{1.76} & \textbf{51s} & 1h40m & 15m32s \\

\midrule
&&&&\multicolumn{3}{c}{Ablation Studies}&&& \\
\midrule

                        & Ablation1 & Ablation2 & Full & Ablation1 & Ablation2 & Full &Ablation1 & Ablation2 & Full \\ 
\cmidrule(lr){2-4} \cmidrule(lr){5-7} \cmidrule(lr){8-10} 
 Mean & \textbf{28.25} & 27.29 & 28.14  & \textbf{1.73} & 1.95 & 1.76  & 17m1s & \textbf{14m50s} & 15m32s \\
\bottomrule
\end{tabular}%
\end{table}

Table \ref{tab1} lists the average PSNR and RMSE of dynamic reconstruction at all time points obtained by different methods and the corresponding computation time. While SuPReMo \cite{huang2024surrogate} achieves the fastest speed, it yields the lowest image quality. Compared with HexPlane \cite{fu2025spatiotemporal}, our proposed method is about six times faster while achieving better image quality. Ablation Study 1 leads to a slight improvement, but remains comparable to results of using a unified DVF. Ablation Study 2 results in a noticeable performance drop.

\begin{figure}[!t]
    \centering
    \includegraphics[width=\textwidth]{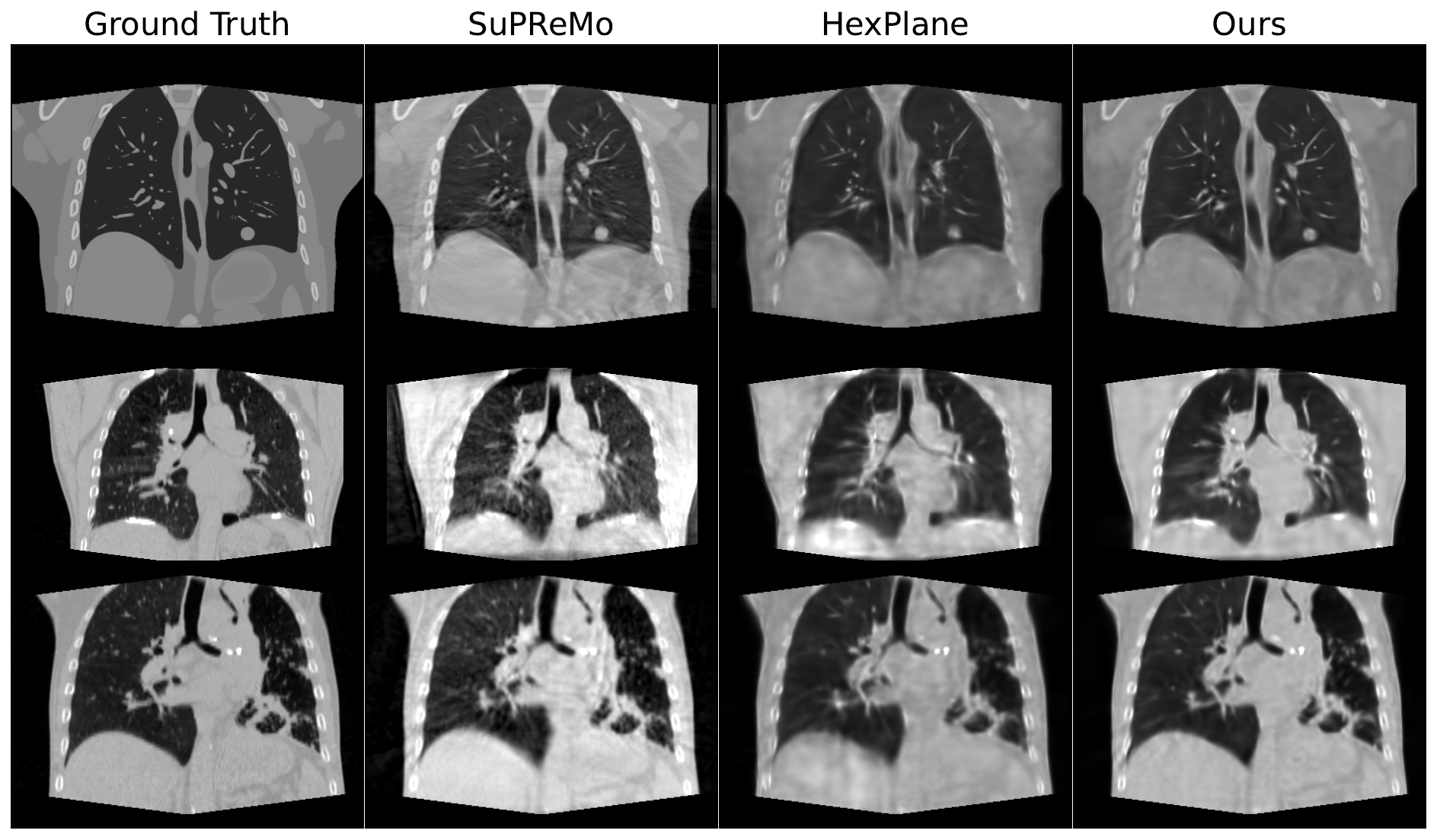}
    \caption{Ground-truth and reconstructed mean position images for case 2 (top), case 3 (middle) and case 5 (bottom). Dynamic range: 0.00 - 0.03 $\mathrm{mm}^{-1}$ }
    \label{fig2}
\end{figure}

Figure \ref{fig2} plots example coronal slices of the reconstructed mean position images for three cases with the lowest, median and highest PSNR values, respectively. Compared with SuPReMo and HexPlane, the proposed method produces sharper diaphragm with less noise and motion artifacts.

\section{Discussion and Conclusion}

Our results demonstrate that the proposed method significantly improves both computational efficiency and image quality in dynamic CBCT reconstruction. Compared to HexPlane, our approach achieves a 6× speedup while producing sharper images with fewer motion artifacts. While SuPReMo is the fastest, it results in the lowest image quality, highlighting the trade-off between speed and accuracy. Qualitative results (Figure \ref{fig2}) further support these findings, showing that our method produces sharper diaphragm boundaries and fewer motion artifacts, even in challenging cases with low PSNR.

Ablation Study 1 assesses the impact of enforcing consistent updates on all the Gaussian parameters, compared to allowing each parameter to evolve independently. It is unsurprising that the decoupled method reconstructs the dynamic images slightly better, as there are more parameters to fit and the Gaussians have more flexibility to match the projection data. The important conclusion is that imposing consistent updates only has a minor impact on reconstruction metrics, but requires fewer parameters, reduces training time, and more importantly enables consistent continuous DVFs to be generated. Ablation Study 2 eliminates the FFD model and fits motion independently for each Gaussian, which significantly degrades performance. This highlights the critical role of spatial regularization in maintaining motion coherence across neighboring Gaussians.

In conclusion, our DI framework and FFD model effectively balances accuracy, efficiency, and robustness for dynamic CBCT reconstruction. Future work will explore physics-based deformation models and further optimization strategies for clinical applications.

\begin{credits}
\subsubsection{\ackname} This study was funded by Elekta Ltd., Crawley and the EPSRC-funded UCL Center for Doctoral Training in Intelligent, Integrated Imaging in Healthcare (i4health) (EP/S021930/1).

\subsubsection{\discintname}
The authors have no competing interests to declare that are relevant to the content of this article.
\end{credits}

%
%
\bibliographystyle{splncs04}
\bibliography{Paper-3968}

\begin{thebibliography}{10}
\providecommand{\url}[1]{\texttt{#1}}
\providecommand{\urlprefix}{URL }
\providecommand{\doi}[1]{https://doi.org/#1}

\bibitem{balik2013evaluation}
Balik, S., Weiss, E., Jan, N., Roman, N., Sleeman, W.C., Fatyga, M., Christensen, G.E., Zhang, C., Murphy, M.J., Lu, J., et~al.: Evaluation of 4-dimensional computed tomography to 4-dimensional cone-beam computed tomography deformable image registration for lung cancer adaptive radiation therapy. International Journal of Radiation Oncology* Biology* Physics  \textbf{86}(2),  372--379 (2013)

\bibitem{bonet1997nonlinear}
Bonet, J., Wood, R.D.: Nonlinear continuum mechanics for finite element analysis. Cambridge university press (1997)

\bibitem{cai2024gaussian}
Cai, J., Yang, Y., Yuan, W., He, Y., Dong, Z., Bo, L., Cheng, H., Chen, Q.: Gaussian-informed continuum for physical property identification and simulation. Advances in Neural Information Processing Systems  (2024)

\bibitem{cao2023hexplane}
Cao, A., Johnson, J.: Hexplane: A fast representation for dynamic scenes. In: Proceedings of the IEEE/CVF Conference on Computer Vision and Pattern Recognition. pp. 130--141 (2023)

\bibitem{de2013image}
De~Los~Santos, J., Popple, R., Agazaryan, N., Bayouth, J.E., Bissonnette, J.P., Bucci, M.K., Dieterich, S., Dong, L., Forster, K.M., Indelicato, D., et~al.: Image guided radiation therapy (igrt) technologies for radiation therapy localization and delivery. International journal of radiation oncology, biology, physics  \textbf{87}(1),  33--45 (2013)

\bibitem{eiben2020consistent}
Eiben, B., Bertholet, J., Menten, M.J., Nill, S., Oelfke, U., McClelland, J.R.: Consistent and invertible deformation vector fields for a breathing anthropomorphic phantom: a post-processing framework for the xcat phantom. Physics in Medicine \& Biology  \textbf{65}(16),  165005 (2020)

\bibitem{eiben2018statistical}
Eiben, B., Tran, E.H., Menten, M.J., Oelfke, U., Hawkes, D.J., McClelland, J.R.: Statistical motion mask and sliding registration. In: Biomedical Image Registration: 8th International Workshop, WBIR 2018, Leiden, The Netherlands, June 28-29, 2018, Proceedings 8. pp. 13--23. Springer (2018)

\bibitem{feng2024gaussian}
Feng, Y., Feng, X., Shang, Y., Jiang, Y., Yu, C., Zong, Z., Shao, T., Wu, H., Zhou, K., Jiang, C., et~al.: Gaussian splashing: Unified particles for versatile motion synthesis and rendering. arXiv preprint arXiv:2401.15318  (2024)

\bibitem{fu2025spatiotemporal}
Fu, Y., Zhang, H., Cai, W., Xie, H., Kuo, L., Cervino, L., Moran, J., Li, X., Li, T.: Spatiotemporal gaussian optimization for 4d cone beam ct reconstruction from sparse projections. arXiv preprint arXiv:2501.04140  (2025)

\bibitem{huang2024resolving}
Huang, Y., Eiben, B., Thielemans, K., McClelland, J.R.: Resolving variable respiratory motion from unsorted 4d computed tomography. In: International Conference on Medical Image Computing and Computer-Assisted Intervention. pp. 588--597. Springer (2024)

\bibitem{huang2024surrogate}
Huang, Y., Thielemans, K., Price, G., McClelland, J.R.: Surrogate-driven respiratory motion model for projection-resolved motion estimation and motion compensated cone-beam ct reconstruction from unsorted projection data. Physics in Medicine \& Biology  \textbf{69}(2),  025020 (2024)

\bibitem{jailin2021projection}
Jailin, C., Roux, S., Sarrut, D., Rit, S.: Projection-based dynamic tomography. Physics in Medicine \& Biology  \textbf{66}(21),  215018 (2021)

\bibitem{kerbl20233d}
Kerbl, B., Kopanas, G., Leimk{\"u}hler, T., Drettakis, G.: 3d gaussian splatting for real-time radiance field rendering. ACM Trans. Graph.  \textbf{42}(4),  139--1 (2023)

\bibitem{kratimenos2024dynmf}
Kratimenos, A., Lei, J., Daniilidis, K.: Dynmf: Neural motion factorization for real-time dynamic view synthesis with 3d gaussian splatting. In: European Conference on Computer Vision. pp. 252--269. Springer (2024)

\bibitem{leng2008streaking}
Leng, S., Zambelli, J., Tolakanahalli, R., Nett, B., Munro, P., Star-Lack, J., Paliwal, B., Chen, G.H.: Streaking artifacts reduction in four-dimensional cone-beam computed tomography. Medical physics  \textbf{35}(10),  4649--4659 (2008)

\bibitem{lin2024phys4dgen}
Lin, J., Wang, Z., Jiang, S., Hou, Y., Jiang, M.: Phys4dgen: A physics-driven framework for controllable and efficient 4d content generation from a single image. arXiv preprint arXiv:2411.16800  (2024)

\bibitem{lin2024gaussian}
Lin, Y., Dai, Z., Zhu, S., Yao, Y.: Gaussian-flow: 4d reconstruction with dynamic 3d gaussian particle. In: Proceedings of the IEEE/CVF Conference on Computer Vision and Pattern Recognition. pp. 21136--21145 (2024)

\bibitem{modat2010fast}
Modat, M., Ridgway, G.R., Taylor, Z.A., Lehmann, M., Barnes, J., Hawkes, D.J., Fox, N.C., Ourselin, S.: Fast free-form deformation using graphics processing units. Computer methods and programs in biomedicine  \textbf{98}(3),  278--284 (2010)

\bibitem{park2021nerfies}
Park, K., Sinha, U., Barron, J.T., Bouaziz, S., Goldman, D.B., Seitz, S.M., Martin-Brualla, R.: Nerfies: Deformable neural radiance fields. In: Proceedings of the IEEE/CVF international conference on computer vision. pp. 5865--5874 (2021)

\bibitem{rit2014reconstruction}
Rit, S., Oliva, M.V., Brousmiche, S., Labarbe, R., Sarrut, D., Sharp, G.C.: The reconstruction toolkit (rtk), an open-source cone-beam ct reconstruction toolkit based on the insight toolkit (itk). In: Journal of Physics: Conference Series. vol.~489, p. 012079. IOP Publishing (2014)

\bibitem{rit2009fly}
Rit, S., Wolthaus, J.W., van Herk, M., Sonke, J.J.: On-the-fly motion-compensated cone-beam ct using an a priori model of the respiratory motion. Medical physics  \textbf{36}(6Part1),  2283--2296 (2009)

\bibitem{rueckert1999nonrigid}
Rueckert, D., Sonoda, L.I., Hayes, C., Hill, D.L., Leach, M.O., Hawkes, D.J.: Nonrigid registration using free-form deformations: application to breast mr images. IEEE transactions on medical imaging  \textbf{18}(8),  712--721 (1999)

\bibitem{segars20104d}
Segars, W.P., Sturgeon, G., Mendonca, S., Grimes, J., Tsui, B.M.: 4d xcat phantom for multimodality imaging research. Medical physics  \textbf{37}(9),  4902--4915 (2010)

\bibitem{sonke2005respiratory}
Sonke, J.J., Zijp, L., Remeijer, P., Van~Herk, M.: Respiratory correlated cone beam ct. Medical physics  \textbf{32}(4),  1176--1186 (2005)

\bibitem{sweeney2012accuracy}
Sweeney, R.A., Seubert, B., Stark, S., Homann, V., M{\"u}ller, G., Flentje, M., Guckenberger, M.: Accuracy and inter-observer variability of 3d versus 4d cone-beam ct based image-guidance in sbrt for lung tumors. Radiation Oncology  \textbf{7}(1), ~1--8 (2012)

\bibitem{xie2024physgaussian}
Xie, T., Zong, Z., Qiu, Y., Li, X., Feng, Y., Yang, Y., Jiang, C.: Physgaussian: Physics-integrated 3d gaussians for generative dynamics. In: Proceedings of the IEEE/CVF Conference on Computer Vision and Pattern Recognition. pp. 4389--4398 (2024)

\bibitem{yang2024deform3dgs}
Yang, S., Li, Q., Shen, D., Gong, B., Dou, Q., Jin, Y.: Deform3dgs: Flexible deformation for fast surgical scene reconstruction with gaussian splatting. In: International Conference on Medical Image Computing and Computer-Assisted Intervention. pp. 132--142. Springer (2024)

\bibitem{r2_gaussian}
Zha, R., Lin, T.J., Cai, Y., Cao, J., Zhang, Y., Li, H.: R$^2$-gaussian: Rectifying radiative gaussian splatting for tomographic reconstruction. In: Advances in Neural Information Processing Systems (NeurIPS) (2024)

\bibitem{zha2022naf}
Zha, R., Zhang, Y., Li, H.: Naf: neural attenuation fields for sparse-view cbct reconstruction. In: International Conference on Medical Image Computing and Computer-Assisted Intervention. pp. 442--452. Springer (2022)

\bibitem{zwicker2002ewa}
Zwicker, M., Pfister, H., Van~Baar, J., Gross, M.: Ewa splatting. IEEE Transactions on Visualization and Computer Graphics  \textbf{8}(3),  223--238 (2002)

\end{thebibliography}

\end{document}